\documentclass{aa}
\usepackage{txfonts}
\usepackage{epsfig}
\usepackage{graphicx}
\usepackage{times}

\def\ltsima{$\; \buildrel < \over \sim \;$}
\def\gtsima{$\; \buildrel > \over \sim \;$}
\def\simlt{\lower.5ex\hbox{\ltsima}}
\def\simgt{\lower.5ex\hbox{\gtsima}}

\begin{document}

\title{$Chandra$ and XMM--$Newton$ observations of Tololo 0109-383}
\author{G. Matt\inst{1}, S. Bianchi\inst{1}, M. Guainazzi\inst{2},  W.N. Brandt\inst{3},
A.C. Fabian\inst{4}, K. Iwasawa\inst{4} \and G.C. Perola\inst{1}}

\offprints{Giorgio Matt\\ \email{matt@fis.uniroma3.it}}

\institute{Dipartimento di Fisica, Universit\`a degli Studi Roma Tre, Italy
\and XMM-Newton Science Operation Center/RSSD-ESA, Villafranca del Castillo,
Spain 
\and
Department of Astronomy and Astrophysics, 525 Davey Laboratory, Pennsylvania State University, University Park, PA 16802, U.S.A.
\and
Institute of Astronomy, University of Cambridge, Madingley Road, Cambridge CB3 0HA, U.K.
}

\date{Received / Accepted}

\authorrunning{G. Matt et al.}

\abstract{ We present and discuss $Chandra$ and XMM-$Newton$ observations of the 
Seyfert 2 galaxy and Compton--thick absorbed source, Tololo~0109-383. The hard 
X--ray emission (i.e. above $\sim$2 keV), is dominated by a spatially unresolved 
reflection component, as already discovered by previous ASCA and BeppoSAX observations.
The soft X--ray emission is partly ($\sim$15\%) extended over about 1 kpc. Below 2 keV,
the spectrum is very steep and two emission lines, probably due to recombination to
He--like ions of oxygen and neon, are clearly present. Combining X--rays and optical
information taken from the literature, we propose an overall scenario for the nuclear
regions of this source.   
}

\maketitle

\section{Introduction}

Tololo~0109--383 (a.k.a. NGC~424) is a remarkable obscured Seyfert galaxy.  
It was classified as a Seyfert 2 by Smith (1975). 
Boisson \& Durret (1986) discovered weak broad  H$\alpha$ and H$\beta$ lines.  
Broad lines
were also observed in polarized light (Moran et al. 2000), the
polarization degree being, after correction for starlight, about 4\%. Murayama et al. (1998) studied the optical
spectrum in detail and, besides confirming the presence of broad  H$\alpha$ and H$\beta$ lines
in direct light, also discovered Fe {\sc ii} emission, and a partially extended 
(about 1 kpc) High Ionization
Nuclear Emission Line Region (HINER), 70\% of which was however unresolved ($\simlt$200 pc).
HST/WFPC2 data showed the presence of a dust lane across the central part of the galaxy (Malkan
et al. 1998), which may help to explain the observed $A_V\sim$1.4 to the NLR 
(Murayama et al. 1998; it
corresponds to N$_H\sim3\times10^{21}$ cm$^{-2}$ for a dust--to--gas ratio equal to 
that of the Galactic ISM).

In X--rays,  Collinge \& Brandt (2000) analysed ASCA data and, based on 
the prominent iron line, the flat spectrum, and the large [{\sc O iii}]/F(2-10 keV) ratio,
argued that the nucleus of Tololo~0109--383 should, rather 
surprisingly given its optical appearance, 
be absorbed by Compton--thick matter. This result was fully confirmed by BeppoSAX
(Matt et al. 2000; Iwasawa et al. 2001), which measured the absorbing 
column to be about 
2$\times10^{24}$ cm$^{-2}$.  The estimated nuclear 2--10 keV luminosity is
about 10$^{43}$ erg s$^{-1}$. The IRAS  colours are quite warm, suggesting
that the IR emission is dominated by dust reprocessing of the nuclear 
radiation (Matt et al. 2000).

In this paper we present $Chandra$ and XMM--$Newton$ observations of 
Tololo~0109--383. The superior perfomances of these satellites permit
us to study the source spectrum down to 0.3 keV and to search for extended
emission. 

Assuming $H_0$=70 km~s$^{-1}$~Mpc$^{-1}$, 
the redshift of the source, $z$=0.0117, corresponds to a
distance of 48.5 Mpc. At this distance, 1\arcsec~ corresponds to 235 pc.

\section{Observations and Data Reduction}

\subsection{Chandra}

$Chandra$ observed Tololo~0109--383 on February 4, 2002, with ACIS--S
in standard configuration and with a 0.8 s frame time, to reduce 
pile--up to negligible
values. After data reduction, performed with CIAO v.2.2.1, the exposure time is 9181 s.

\subsection{XMM--Newton}

XMM--$Newton$ observed the source on December 12, 2001. Both EPIC cameras
were operated in full--frame mode, as at the flux level expected for the source
pile--up is negligible.  The data were 
reduced with {\sc SAS 5.3.3}, and using a Calibration Index File 
generated at the time of the analysis, August 15, 2002.
Only events corresponding to patterns 0-4
and 0-12 were used for the p-n and MOS, respectively. 
Events from the two MOS cameras were merged to obtain a single event file. 
The background in both instruments remained constant during the observation.
After data reduction, the exposure times are 4525 s for the p-n, and 7538 s for the MOS.

\section{Data Analysis}

The data were binned in order to have at least 25 counts per bin, to ensure applicability 
of $\chi^2$ statistics, and to oversample the energy resolution by a factor
3. They have been analyzed with {\sc xspec.v11.1}. Errors correspond
to the 90\% confidence level for one interesting parameter ($\Delta\chi^2$=2.7).

\subsection{Chandra}

\subsubsection{Spatial analysis}

\begin{figure}
\epsfig{figure=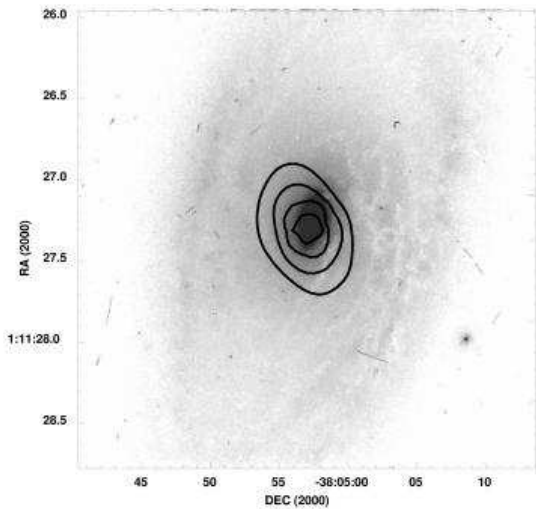,width=100mm}
\caption{ The $Chandra$ 0.3-7 keV isophotes superimposed on the HST image.}
\label{chimage}
\end{figure}

\begin{figure}
\epsfig{file=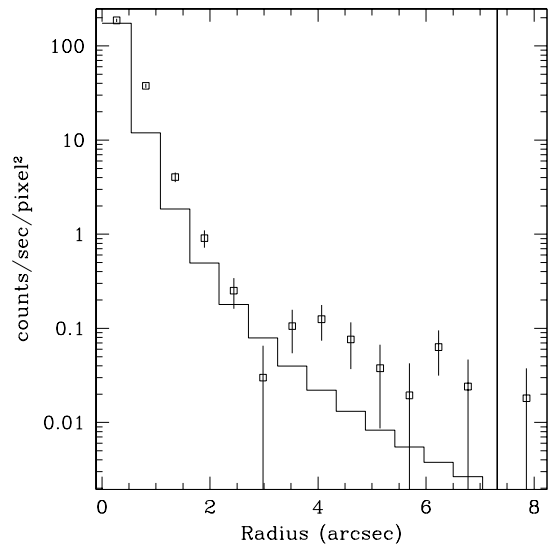,height=80mm}
\caption{ The radial dependence of the source counts (squares) in the $Chandra$/ACIS-S
0.3-7 keV image, compared with the PSF (histogram, obtained with the {\sc mkpsf} tool
in the software package CIAO.)}  
\label{tol_ext}
\end{figure}

The image is dominated by a point--like source coincident, within
the $Chandra$ angular resolution, with the optical nucleus (see 
Fig.~\ref{chimage}). 
However, emission extended over about 5\arcsec~ is also apparent
(Fig~\ref{tol_ext}). This emission is about 13\% (17\%) of the total flux
in the 0.3-10 (0.3-2) keV 
energy band, and it is somewhat asymmetric (Fig.~\ref{chimage}).

\subsubsection{Spectral analysis}

The $Chandra$ spectra were fitted in the 0.3-7 keV energy range (at higher energies no
source emission is detected).

We first analysed the spectrum of the extended emission alone (taken from an annulus
with inner and outer radii of 1\arcsec~ and 5\arcsec~, respectively). The count rate of this region,
over the adopted 0.3-7 keV energy range, is 0.017 cts/s. The spectrum is quite soft,
and it is reasonably 
well fitted either by an absorbed power law with $\Gamma$=3.0$^{+0.7}_{-0.5}$
and $N_H$=1.4$^{+1.2}_{-0.5}\times10^{21}$ cm$^{-2}$ 
 ($\chi^2_{\rm r}$=1.0/6 d.o.f.), or by a thermal plasma 
spectrum (model {\sc mekal}; ($\chi^2_{\rm r}$=1.2/5 d.o.f.)
with $kT$=1.2$^{+1.2}_{-0.7}$ keV; for the latter model, only 
upper limits of 0.19 solar and
of 0.9$\times10^{21}$ cm$^{-2}$ can be put on the metal abundance and the absorber
column density, respectively. Given the implausibly low metal abundance, in the following
we will adopt the power law model (even if, of course, the low metal abundance
may derive from a too simple thermal plasma model; however, statistics 
is not good enough to test more complex models). In this case, absorption in excess
of the Galactic one (1.8$\times10^{20}$ cm$^{-2}$, Dickey \& Lockman 1990) is required,
and may be due to the dust lane observed by HST (Malkan et al. 1998). The value of the
$N_H$ is broadly in agreement with the reddening to the NLR, which is $A_V$=1.4
(Murayama et al. 1998).

\begin{figure}
\epsfig{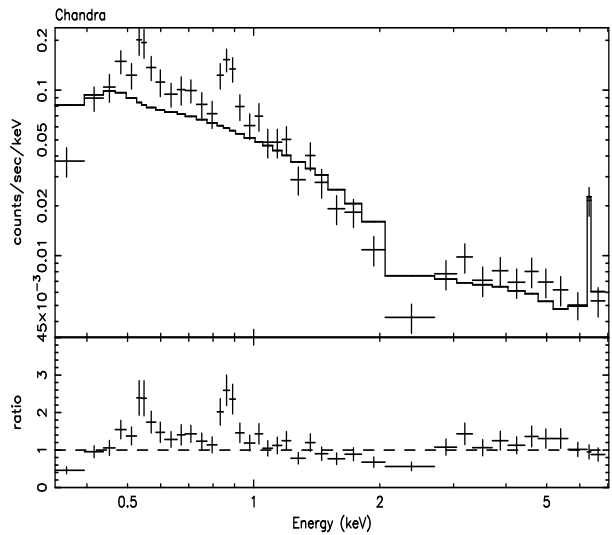}
\caption{ The $Chandra$/ACIS-S spectrum when fitted with  the continuum
components as observed by BeppoSAX, plus the 6.4 keV iron line (see the text).}
\label{ch_badsp}
\end{figure}

\begin{figure}
\epsfig{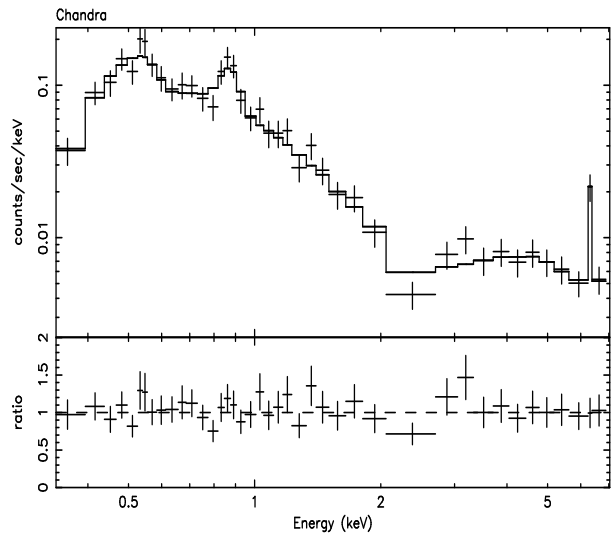}
\caption{  The $Chandra$/ACIS-S spectrum and best fit model. }
\label{ch_goodsp}
\end{figure}

We then analyzed the spectrum extracted from a circular cell
with a radius of 5\arcsec~ (count rate of 0.14 cts/s). 
(We choose to analyze the unresolved+extended
spectrum, instead of the nuclear spectrum 
alone, to make easier the comparison with the XMM--$Newton$
spectrum, as well as with spectra from previous satellites. In any case,
unresolved emission dominates at all energies.) Following Iwasawa et al. (2001)
we adopted a model composed of: a nuclear power law absorbed by a cold screen of 
2$\times10^{24}$ cm$^{-2}$; a cold reflection component ({\sc pexrav} model;
the illuminating power law index
has been initially fixed to 2, given the limited statistics available); a 6.4 keV narrow iron line;
a soft X-ray emission component, parameterized by a power law; and
Galactic absorption. The fit
is completely unacceptable ($\chi^2_{\rm r}$=4.4/36 d.o.f.;
 see Fig.~\ref{ch_badsp}), partly due to two emission 
features at about 0.55 and 0.9 keV (the latter was 
indeed already found by Iwasawa et
al 2001 in the ASCA spectrum).  The inclusion of these two lines improves the fit
significantly; however, the $\chi^2$ is still unacceptable ($\chi^2_{\rm r}$=2.3/32 d.o.f.). Adding a 
thermal plasma component further improves the quality of the fit 
($\chi^2_{\rm r}$=1.68/29 d.o.f.), which however remains poor.
A much better, and fully acceptable, fit ($\chi^2_{\rm r}$=0.88/31 d.o.f.) is instead
obtained by allowing the
cold absorption, covering all components, to be larger than 
the Galactic value. Interestingly, the best fit value, $\sim3\times10^{21}$ 
cm$^{-2}$, is  what is expected from the optical extinction.
No significant improvement is found
after adding either a He--like or a H--like iron line. 
The best--fit values are summarized in Table 1, and the spectrum is shown in Fig.~\ref{ch_goodsp}.
As the power law index of the soft component is much larger than 2, the value we assumed
for the hard X-ray component (both transmitted and reflected), we tried to link the two
values to each other. The resulting value of $\Gamma$ is 2.9, but 
the fit is unacceptable ($\chi^2_{\rm r}$=1.96/31 d.o.f.).
A fit with a thermal plasma model instead of the soft power law plus two
lines also fails to give an acceptable fit ($\chi^2_{\rm r}$=1.8/34 d.o.f) mainly because the two lines
remain unfitted. Adding instead either the thermal model or a black body 
to the best fit spectrum, no further improvement is found. 

To understand whether the two soft X--ray lines can originate in the extended emission, 
we refitted its spectrum adding the two lines in turn (with the energies 
fixed to 0.57 and 
0.92 keV, respectively; see Sect.~4). Neither line is required, the upper limits 
being 30$\times$10$^{-5}$ and 10$^{-5}$ ph cm$^{-2}$ s$^{-1}$, respectively. The latter
is inconsistent with the value obtained from the total spectrum. Assuming that
the two lines are emitted by the same material, we can conclude that at least a large
part of them, and possibly all, is associated with the unresolved component. 

The 2--10 (0.5--10) keV observed flux is 1.6(1.8)$\times10^{-12}$ erg cm$^{-2}$ s$^{-1}$,
corresponding to an observed luminosity at source of 4.8(5.4)$\times10^{41}$ erg s$^{-1}$.

\subsection{XMM--$Newton$}

At the spatial resolution of XMM--$Newton$, the source is point like. The p-n
(MOS) spectrum was extracted using a circular source cell with
a radius of 40\arcsec~ (45\arcsec~) and fitted in the 0.5-10
(0.3-10) keV energy range, where the instrument calibration is reliable. 
The MOS spectra were combined together, once we verified that the
spectral results from the individual units were consistent
within the statistical
uncertainties. In this energy range, the count rate is 0.27 (0.18) cts/s.
The flux of the source and the exposure time are too low to give a
significant detection in the RGS. 

\begin{figure}
\epsfig{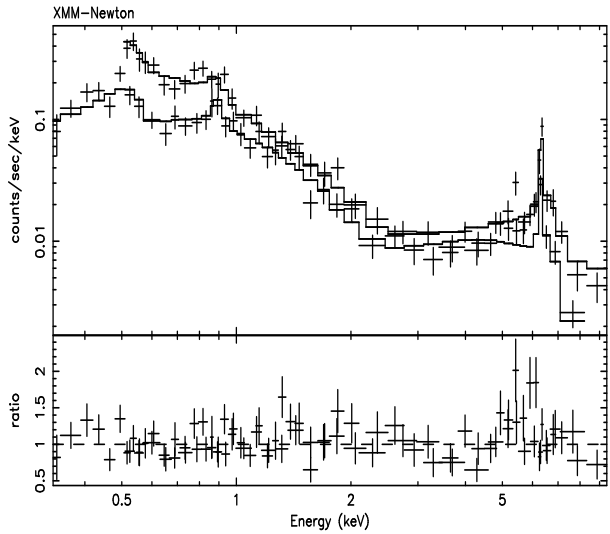}
\caption{ The XMM--$Newton$ EPIC p-n and MOS spectra and best--fit model.}
\label{xmm_goodsp}
\end{figure}

\begin{figure}
\epsfig{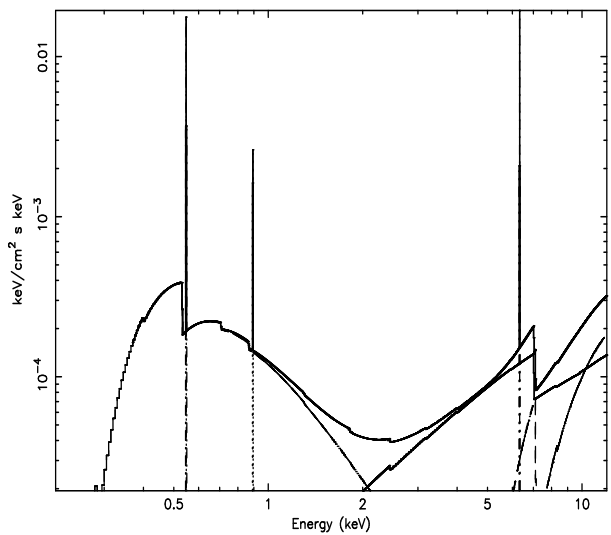}
\caption{ The best fit model for the  XMM--$Newton$ spectra. Starting from the right, the
following components are present: the heavily absorbed nuclear continuum; the cold
reflection component; the 6.4 keV iron line; the soft power law; the neon He-like
line; the oxygen He-like line. The upper curve is the sum of all components. }
\label{model}
\end{figure}

We fitted the XMM--$Newton$ EPIC spectra
with the same model adopted for the Chandra spectrum. The fit is
acceptable ($\chi^2_{\rm r}$=1.22/78 d.o.f.; see Fig.~\ref{xmm_goodsp} and Table~1). Again, 
 neither a He--like nor a H--like iron line is required by
the data. Adding the iron K$\beta$ component at 7.06 keV, no significant 
improvement ($\Delta\chi^2=0.5$) is found. The best--fit value for its flux, 
however, is about 17\% that of the K$\alpha$ component, roughly as expected.
The best--fit model is shown in Fig.~\ref{model}. Again, substituting the soft power law
plus lines with a thermal plasma model gives a significantly worse fit ($\chi^2_{\rm r}$=1.40/81),
besides an unacceptably low ($<$0.02) metal abundance. The addition of either a thermal
plasma or black body component to the best fit model of Table~1
results in no statistical improvement. 

In Fig.~\ref{xmm_goodsp}, some residuals 
around the iron K$\alpha$ line are apparent. We therefore allowed both the line energy and
width to vary. No significant improvement is however found 
($\Delta\chi^2$=0.8 with two less d.o.f.). 
The best fit line energy is 6.38$\pm$0.04 keV, and the upper limit 
to $\sigma$ is 90 eV.
 
Finally, the 2--10 keV (0.5--10) observed 
flux is 1.6(1.9)$\times10^{-12}$ erg cm$^{-2}$ s$^{-1}$, in agreement 
with the $Chandra$ one (as well as with ASCA and BeppoSAX fluxes, Iwasawa
et al. 2001).

\begin{table}[t]

\caption{$Chandra$ and XMM--$Newton$ best fit parameters (see text for
details). All lines are $\delta$-functions. $N_H$ and $N_{\rm H, nucl}$
refer to the cold absorbers in front of all components and of the nuclear continuum alone,
respectively.}
\begin{center}
\begin{tabular}{lcc}
\hline 
\hline
~ & ~ & ~\cr
Parameter & $Chandra$ & XMM--Newton \cr
~ & ~ & ~\cr
\hline
~ & ~ & ~\cr
$N_H$~(10$^{21}$~cm$^{-2})$ & 2.7$^{+0.7}_{-0.6}$ & 1.8$^{+0.3}_{-0.5}$ \cr
$\Gamma_{\rm soft}$ & 4.1$^{+0.4}_{-0.5}$ & 3.9$^{+0.2}_{-0.2}$ \cr
$E_{l,1}$~(keV) & 0.57$^{+0.02}_{-0.04}$  & 0.55$^{+0.04}_{-0.03}$ \cr
$F_{l,1}$~(10$^{-5}$~ph~cm$^{-2}$~s$^{-1}$) & 30$^{+31}_{-12}$ & 18$^{+29}_{-12}$ \cr
$EW_{l,1}$~(eV)  & 150 & 90 \cr
$E_{l,2}$~(keV) & 0.89$^{+0.03}_{-0.02}$  & 0.90$^{+0.03}_{-0.03}$ \cr
$F_{l,2}$~(10$^{-5}$~ph~cm$^{-2}$~s$^{-1}$) & 3.0$^{+1.7}_{-1.1}$ &  1.5$^{+0.7}_{-0.6}$ \cr
$EW_{l,2}$~(eV)  & 100 & 50 \cr
$E_{l,3}$~(keV) & 6.4 (fixed) & 6.4 (fixed) \cr
$F_{l,3}$~(10$^{-5}$~ph~cm$^{-2}$~s$^{-1}$) & 2.4$^{+0.8}_{-0.9}$ & 1.5$^{+0.4}_{-0.3}$ \cr
$EW_{l,3}$~(eV)  & 960  & 790 \cr
$\Gamma_{\rm hard}$ & 2 (fixed) & 2 (fixed) \cr
$N_{\rm H, nuc}$~(10$^{24}$~cm$^{-2})$ &  2 (fixed) & 2 (fixed) \cr
$\chi^2_{\rm r}$/d.o.f & 0.82/31 & 1.22/78 \cr
~ & ~ & ~\cr
\hline
\hline
\end{tabular}
\end{center}
\end{table}

\section{Discussion}

\subsection{The soft X--ray emission }

The soft X--ray emission of Tololo~0109-383 is partly (about 15\%) extended
over a scale of $\sim$5\arcsec~, interestingly similar to the scale over which 
part of the HINER is extended (Murayama et al. 1998). 
However, the ionization structure
of the HINER is too low to reflect soft X--rays efficiently, and therefore
the two regions are probably not associated with each other.  The quality of the spectrum
of the extended emission is not good enough to allow for a detailed spectral
analysis. The spectrum is, in any case, consistent with that of the unresolved soft
X--ray emission, suggesting a possible common origin as reflection 
of the nuclear radiation from ionized matter.

Two emission lines, partly if not entirely coming from the unresolved component, are
also clearly present. One of them, at about 0.9 keV, was already observed by Iwasawa
et al. (2001) in the ASCA spectrum. Its energy is consistent with 
either the O~{\sc viii}
recombination continuum (0.87 keV) or with the Ne~{\sc ix} (0.92 keV) recombination line.
A contribution from iron L lines is also possible. 
The other line, at $\sim$0.55 keV, is most naturally explained as a O~{\sc vii}
recombination line (0.57 keV). If the two lines originate from the same 
material, the $\sim$0.9 keV line is therefore more likely due to Ne~{\sc ix}. 
It is worth noting that the emitting region cannot be associated with the 
HINER, as the ionization of the latter is much lower. 

A possible site of origin for these two lines is the same reflector responsible for
the iron K$\alpha$ line, if it is mildly ionized, similar to what was
found by Bianchi et al. (2001) for the Circinus Galaxy. They found that an
ionization structure in which He--like ions of light elements
may coexist with iron less than {\sc xvii} is possible. 
We therefore fitted both the $Chandra$ and XMM--$Newton$
spectra with the {\sc pexriv} model for ionized reflection instead of the {\sc pexrav}
model, with the ionization parameter fixed to the same value found for the Circinus 
Galaxy (Bianchi et al. 2001). In both cases the fit is significantly
worse ($\chi^2_{\rm r}$=1.02/31 d.o.f and 1.55/78 d.o.f. for $Chandra$ and XMM--$Newton$, respectively),
and the soft power law is steeper than before ($\Gamma\sim$5). The equivalent widths of the
two lines with respect to the 
mildly ionized reflector ($\sim$3 and 0.9 keV, respectively) are however
reasonable (e.g. Matt et al. 1996). 
Therefore, we consider this model as less favoured but still possible.
 
The power law photon index of the soft component is large, i.e. around 4.
 Assuming that this
component is due to reflection of the nuclear radiation, and neglecting self-absorption
effects, this would imply a soft (i.e. below about 2 keV) X--ray nuclear
emission more typical of Narrow--Line Seyfert 1s (Boller et al. 1996) than of
classical Seyferts. 
Indeed, Boisson \& Durret (1986) measured for the broad lines a FWHM
of 1800 km~s$^{-1}$, slightly lower than the value of 2000 km~s$^{-1}$ conventionally adopted as the
boundary between Narrow--Line and classical Seyfert 1s.
However, Murayama et al. (1998) measured a FWHM of about 3800 km~s$^{-1}$
 for the
broad H$\beta$ component (but only 1500 km~s$^{-1}$ for the H$\alpha$), and Moran et al. (2000)
found even larger wings in the lines observed in polarized light. In this case, 
Tololo~0109-383 would probably be the most X--ray soft among classical Seyferts.

At least part of the soft excess may however be due to several unresolved emission lines
from photoionized plasma, similarly to what found in NGC~1068 (Kinkhabwala et al. 2002),
weakening the case for a very steep continuum. Unfortunately, the source is too faint
for the RGS to be profitably used, and we cannot check directly this hypothesis.

\subsection{The absorption structure}

Combining previous ASCA and BeppoSAX results (Collinge \& Brandt 2000; Iwasawa et al. 2001) 
with the $Chandra$ and XMM--$Newton$ observations discussed here, we can conclude that
at least two X--ray absorbers are present: one Compton--thick, obscuring the 
nucleus on a small scale, 
the other Compton--thin, obscuring the soft X--ray emission and possibly the extended emission.
The latter may be associated with the dust lanes observed by HST to obscure the central
part of the galaxy (Malkan et al. 1998). This provides one more piece of
evidence in favour of the
co--existence of both Compton--thin and Compton--thick matter in the circumnuclear
regions of Seyfert galaxies (see e.g. Matt \& Guainazzi 2002 and references therein).

The problem here is that the optical broad lines appears to be seen 
through the dust lane rather than the Compton--thick X-ray absorber.
Recently, it has become clear that a fraction of type 1 
nuclei are absorbed in X--rays (e.g. Maiolino et al. 2001a; Fiore et al. 2001)
and, more generally, that X--ray absorption column densities are often 
much larger
than would be expected from the amount of optical extinction
(e.g. Granato et al. 1997). While a fraction
of hard X-ray spectrum, type 1 AGN may be explained in terms 
of a temporary switching--off
of the nucleus, which makes them for a while 
reflection--dominated and so apparently Compton--thick
absorbed (Matt et al. 2002), this is certainly not the case for Tololo~0109-383,
whose nucleus is definitely absorbed by Compton--thick matter (Iwasawa et al. 2001).
Let us call this matter, for simplicity, the `torus'.

There are several possible solutions to this problem (see Maiolino et al. 2001b for a discussion). 

First of all, the Broad Line Region (BLR)
may be located outside the torus. However, the typical BLR
size, from reverberation mapping studies, is usually of the order of light-days
or of light-weeks, at least for a moderately luminous source as Tololo~0109-383,
 while the inner surface of the torus is expected to have a size of a 
fraction of a pc or more
(e.g. Bianchi et al. 2001 and references therein). 
Moreover, the ratio between the fluxes of the broad and narrow components 
is very low (less than 1, Murayama et al. 1998).
Therefore, even though
we cannot exclude that the size of either the BLR or the torus, or both, 
are different than usual in this object (we do not have any direct measurement 
of them) this possibility seems unlikely. 

Another possibility is that the dust--to--gas ratio of the absorber is very low, due
to dust sublimation (Granato et al. 1997). However, the  H$\alpha$/H$\beta$ 
ratio is similar for the narrow and broad components, suggesting that they are observed through
the same absorber, which is probably the dust lane observed by HST. 
Therefore, dust sublimation must be almost complete, to avoid further 
extinction of the
broad components, and the low fluxes of the broad components would remain unexplained. 
An alternative solution, proposed by
Maiolino et al. (2001b), is that the sizes of the dust grains are much larger
than that in the ISM of our own Galaxy, changing dramatically the extinction curves. This
solution has the merit of explaining qualitatively the low fluxes 
of the broad line components, 
but the expected extinction, given the X--ray measured column density, would be too high to
allow them to be observed at all (see the figures in Maiolino et al. 2001b). 

All these problems can be avoided if the broad lines are actually seen in 
reflected, rather than direct, light. Indeed, the 
broad lines in this object are best 
seen in polarized light (Moran et al. 2000) and therefore at least part
of them must be reflected. It is possible that the reflecting medium is the same
responsible for the  soft X--ray excess and the O~{\sc vii} and Ne~{\sc ix} 
lines we observe in the $Chandra$ and XMM--$Newton$ spectra. One may wonder
why in this source the reflected light would be so intense to permit the
detection of broad lines in direct light, despite the fact that the 
polarization degree is by no means exceptional. This is possible if the
covering factor and optical depth of the reflecting matter are large
enough; it must be recalled that a large covering factor implies a low
polarization degree, for obvious geometrical reasons. Unfortunately,
this hypothesis cannot be readily tested in X--rays because the evidence
of a spectral break in the nuclear spectrum makes it difficult to estimate
the nuclear--to--scattered flux ratio.

\section{Summary}

To summarize, the $Chandra$ and XMM--$Newton$ observations, together with previous
X--ray and optical observations, suggest the following scenario for Tololo~0109-383.
The nucleus is absorbed by Compton--thick material, the nuclear radiation being reflected
by: a)  cold material (probably the inner wall of the torus) giving rise to the Compton
reflection component and the iron K$\alpha$ line (and, by reprocessing, to the 
infrared emission, Matt et al. 2000 and Iwasawa et al. 2001); b) ionized matter, responsible 
for the soft X--ray excess, and the oxygen and neon He--like lines. This ionized matter may
coincide with that reflecting and polarizing 
the otherwise 
obscured BLR. Further material, partly spatially resolved, is responsible for the HINER;
the size of the spatially extended HINER emission ($\sim$1 kpc) is similar to that of the 
extended soft X--ray emission, but probably the two regions are not associated
with each other. Finally, all these components are seen through a dust lane, 
responsible for the Balmer decrement and the absorption of the 
soft X--ray emission.

\section*{Acknowledgements} 

GM and SB acknowledge ASI and MIUR (under
grant {\sc cofin-00-02-36}) for financial support. WNB acknowledges
the Chandra X-ray Center grant GO2-3123X
and the NASA LTSA grant NAG5-8107.
This research has made use of the NASA/IPAC Extragalactic Database (NED) which is operated
by the Jet Propulsion Laboratory, California Institute of Technology, under contract
 with the National Aeronautics and Space Administration. This paper is partly
based on observations obtained with XMM-Newton, an ESA science
mission with instruments and contributions directly
funded by ESA Member States and the USA (NASA).

{}

\end{document}